\documentclass[aps,12pt,a4paper,onecolumn,showpacs,showkeys,superscriptaddress]{revtex4}
\usepackage{amsfonts}
\usepackage{amsmath}
\usepackage{amssymb}
\usepackage{graphicx}
\usepackage{subfigure}
\begin{document}
\title{The Coulomb Phase Shift Revisited}

\author{J.\ C.\ A.\ Barata}
\email{jbarata@if.usp.br}
\affiliation{Instituto de F\'{i}sica, Universidade de S\~{a}o Paulo,
  C.P.\ 66318, 05314-970 S\~{a}o Paulo, SP, Brazil}
\author{L.\ F.\ Canto}
\email{canto@if.ufrj.br}
\affiliation{Instituto de F\'{\i}sica, Universidade Federal do Rio de Janeiro, C.P.\
68528, 21941-972 Rio de Janeiro, RJ, Brazil\\ and\\
 Centro Brasileiro de Pesquisas F\'{i}sicas (CBPF), 
Rua Xavier Sigaud, 150, 22290-180 Rio de Janeiro, RJ, Brazil}
\author{M.\ S.\ Hussein}
\email{hussein@if.usp.br}
\affiliation{Instituto de F\'{i}sica, Universidade de S\~{a}o Paulo,
  C.P.\ 66318, 05314-970 S\~{a}o Paulo, SP, Brazil}

\begin{abstract}
  We investigate the Coulomb phase shift, and derive and analyze new 
  and more precise analytical formulae.  We consider
  next to leading order terms to the Stirling approximation, and show
  that they are important at small values of the angular momentum $l$
  and other regimes. We employ the uniform approximation. The use of our 
  expressions in low energy scattering of charged particles is discussed 
  and some comparisons are made with other approximation methods.
\end{abstract}
 
\maketitle

\section{Introduction}

The customary procedure to deal with charged particle scattering is to
partial wave the amplitude and identify the Coulomb phase shift from
which scattering information can be obtained, by adding to the Coulomb
amplitude, the contribution from whatever other short-range potential.
In many applications, the asymptotic form of large angular momentum is
employed for the Coulomb phase shift. In this paper we revisit
the derivation of this asymptotic and derive next to leading order
correction to the usual WKB from.

First, we recall the form of the Coulomb phase shift \cite{Brink85}
\begin{equation}
e^{2i\sigma_l} = \frac{\Gamma(1+l+i\eta)}{\Gamma(1+l-i\eta)},
\label{sigma-gamma}
\end{equation}
where $\Gamma$ is Euler's gamma function, $l$, a non-negative integer,
is the angular momentum and the real parameter $\eta$ is the so-called
Sommerfeld parameter, which is inversely proportional to the square root of the scattering
energy. 

In this paper we will present simple formal proofs of some asymptotic
approximations for the gamma function, like Stirling series and
Gudermann series, and we will discuss several methods for
approximately computing the phase shift, Eq. (\ref{sigma-gamma}), from these
asymptotic approximations in various regimes. We will compare these
results with the corresponding ones obtained from other methods for
computing phase shifts, like the WKB and the eikonal
approximations. We will also present proofs of some \underline{exact}
relations for the phase shifts $\sigma_l$ which are more or less known
in the literature. For instance, using Eq. (\ref{sigma-gamma}) we will
easily show that
\begin{equation}
\sigma_l 
=
\sigma_0 + \sum_{m=1}^l\tan^{-1}\left( \frac{\eta}{m} \right),
\label{Formula-exata-para-sigma-l-I}
\end{equation}
and that 
\begin{equation}
\sigma_0 = 
-\gamma\eta - \sum_{m=0}^\infty \left[\tan^{-1}\left(
    \frac{\eta}{m+1}\right) - \frac{\eta}{m+1}\right] ,
\label{Formula-exata-para-sigma-0-I}
\end{equation}
valid for all $\eta\in\mathbb{R}$, from which we obtain for $|\eta|<1$
the power series representation
\begin{equation}
\sigma_0 = 
-\gamma\eta - \sum_{k=1}^{\infty}\frac{(-1)^k\zeta(2k+1)}{2k+1} \eta^{2k+1},
\label{Formula-exata-para-sigma-0-II}
\end{equation}
where $\zeta$ is Riemann's zeta function and $\gamma$ is Euler's constant.

Using our asymptotic expansions we show that 
\begin{eqnarray}
\sigma_0^{(1)} 
=
\frac{1}{2}\tan^{-1}\left( \eta \right)
+ \eta\left( \ln\left(\sqrt{ 1 + \eta^2 }\right) -1\right)
- \frac{\eta} {12\,\left(1+\eta^2\right)} 
\end{eqnarray}
is an excellent approximation for the exact expression
(\ref{Formula-exata-para-sigma-0-I}). We also show numerically that
this approximation is very good even for small values of $\eta$.
More generally, we show that 
\begin{eqnarray}
\sigma_l^{(1)}
 = 
\left( l+ \frac{1}{2}\right)\tan^{-1}\left( \frac{\eta}{l+1} \right)
+ \eta\bigg( \ln\Big(\sqrt{(l+1)^2  + \eta^2} \Big) -1\bigg)
- \frac{\eta} {12\,\big((l+1)^2+\eta^2\big)} 
\end{eqnarray}
approximates (\ref{Formula-exata-para-sigma-l-I}) very well for
$\sqrt{(l+1)^2+\eta^2}$ ``large''.

\section{Asymptotic properties of the gamma function}

We start with the well-known integral representation for Euler's gamma
function, valid for $\mbox{Re}(z)>-1$ \cite{AS},
\begin{equation}
\Gamma(1+z)=\int_{0}^{\infty}dt\, t^{z}\, e^{-t} .
\label{EulerIntegral}
\end{equation}
For (\ref{sigma-gamma}) we will take $z=l +i\eta$. Writing $t^{z} = e^{z\ln{t}}$, we can use
the saddle point method to evaluate the integral
\begin{equation}
\Gamma(1+z)=\int_{0}^{\infty}dt\, \, e^{-t + z\ln(t)} .\label{gamma-int}
\end{equation}
Expanding the exponent of the integrand around its extremum, at $t_0 =z$ and keeping
terms up to second order,
\begin{equation}
-t + z\ln(t) \simeq -z + z\ln(z) - \frac{1}{2z}(t-z)^2,
\end{equation}
Eq.~(\ref{gamma-int})  becomes 
\begin{equation}
\Gamma(1+z)\simeq e^{-z} z^{z}\, \int_{0}^{\infty}dt\,  e^{-(t-z)^2/2z}.
\end{equation}
For $\mbox{Re}(z)=l \gg 1$ and $\mbox{Im}(z)=\eta>0$, the Gaussian
integral can be easily evaluated as
\begin{equation}
\int_{0}^{\infty}~dt\, e^{-(t-z)^2/2z}=
\int_{-\infty}^{\infty}~dt\, e^{-(t-z)^2/2z}= \sqrt{2\pi z}
\end{equation} 
and using this result in the previous equation, we obtain
\begin{equation}
\Gamma(1+z) \simeq \sqrt{2\pi}\, e^{-z}\, z^{z+1/2}.
\label{Gamma-large-l}
\end{equation}
This is the well-known Stirling's approximation for the gamma
function, whose validity can be established for the whole complex
plane, except the non-positive real axis, i.e., in
$\mathbb{C}^-=\mathbb{C}\setminus(-\infty, \; 0]$, and for ``large''
values of $|z|$. See, e.g., Ref.\ \cite{Remmert}.

On the real line, Stirling's approximation (\ref{Gamma-large-l}) has a
long history having been first presented by de Moivre around 1730,
who found an approximation for $n!$, valid for ``large'' $n$, in the
form $n! \simeq K \, n^{n + \frac{1}{2}} e^{-n}$, for some
unspecified constant $K$. In the same year, Stirling proved that $K=
\sqrt{2\pi}$ using Wallis product formula for $\pi$. This
approximation became known as Stirling's approximation for $n!$ and
quite soon generalizations for Euler's gamma function on the positive
real line became available. Stirling's approximation is very useful in
Statistics and Probability Theory because if offers a very good
approximation for ``large'' factorials. For small values of $n$,
however, correcting factors are in order. Stirling found corrections
for the approximation that bears his name for $n!$ in terms of an
asymptotic (but not convergent!) series in $1/n$ that became known as
Stirling's series. Further analysis of those corrections for Euler's
gamma function $\Gamma(z)$, valid on the complex half-plane
$\mbox{Re}(z)>0$, have been performed by Binet in 1839
(Ref.~\cite{Binet}). Another important contribution was made by
Gudermann in 1845 (Ref.~\cite{Gudermann}), who found another
expression for the correcting factors in terms of a convergent
expansion of another kind. The most important contribution to the
study of corrections to Stirling's approximation on the complex plane
was the work of Stieltjes, dated of 1889 (Ref.~\cite{Stieltjes}), who
generalized Stirling's approximation and Gudermann's corrections to
the whole complex plane, excluding the negative real axis (i.e., to
$\mathbb{C}^-$). For a more detailed account of the developments on
the complex plane, see \cite{Remmert} and
\cite{Remmert-on-Wielandt}. For some recent contributions to the
correcting factors to factorials, see \cite{HRobbins} and
\cite{Mansour}. This last reference contains a list of historical
results on corrections on Stirling's approximation.

Let us briefly describe the ideas behind Gudermann's corrections and
present Stirling's series. Since Euler's gamma function satisfies
$\Gamma(z+1)=z\Gamma(z)$, we get from (\ref{Gamma-large-l}) the
approximation
\begin{equation}
\Gamma(z) \simeq \sqrt{2\pi}\, e^{-z}\, z^{z-1/2}.
\label{Gamma-large-l-A}
\end{equation}
However, (\ref{Gamma-large-l-A}) contrasts with the expression
obtained from (\ref{Gamma-large-l}) itself by replacing $z$ by $z-1$:
\begin{equation}
\Gamma(z) \simeq \sqrt{2\pi}\, e^{-(z-1)}\, (z-1)^{z-1/2}.
\label{Gamma-large-l-B}
\end{equation}
Since both expressions (\ref{Gamma-large-l-A}) and
(\ref{Gamma-large-l-B}) are only valid for $|z|$ very large, there is
no practical difference between them. Nevertheless, one can better deal with
this situation by seeking an \underline{exact} representation for
$\Gamma$ in the whole region $\mathbb{C}^-$ (and not only for
``large'' $|z|$) in the form
\begin{equation}
\Gamma(z) = \sqrt{2\pi}\, e^{-z}\, z^{z-1/2} e^{\mu(z)},
\label{Gudermann-1}
\end{equation}
and fixing the correction factor $e^{\mu(z)}$ by imposing the relation 
 $\Gamma(z+1)=z\Gamma(z)$. A simple computation reveals that this
 condition implies that $\mu$ has to satisfy the functional equation
\begin{equation}
  \mu(z) - \mu(z+1)  = 
 \left( z + \frac{1}{2} \right) \ln\left( 1 + \frac{1}{z} \right) -1 .
\label{functional-equation-for-mu}
\end{equation}
Moreover, the validity of (\ref{Gamma-large-l-A}) for ``large'' $|z|$
leads to the condition $\lim_{|z|\to\infty}\mu(z)=0$. This allows to a
solution for (\ref{functional-equation-for-mu}). Indeed, it follows
immediately from (\ref{functional-equation-for-mu}) that for any
positive integer $n$ one has
\begin{equation}
\mu(z) - \mu(z+n)  = 
 \sum_{m=0}^{n-1}
\left[
\left( z + m + \frac{1}{2} \right) \ln\left( 1 + \frac{1}{z+m} \right) -1 \right].
\end{equation}
Hence, the condition $\lim_{|z|\to\infty}\mu(z)=0$ implies, in
particular, that $\lim_{n\to\infty}\mu(z+n)=0$ and we get
\begin{equation}
\mu(z) 
\; = \; 
\sum_{m=0}^\infty 
\left[ 
  \left(z+m+\frac{1}{2}\right) \ln\left( 1 + \frac{1}{z+m}\right) -1
\right] .
\label{Gudermann-2}
\end{equation}
This series is known as Gudermann series. According to
\cite{Remmert}, it was first obtained by that author in 1845
\cite{Gudermann} for real and positive $z$ and the generalization for
$z\in\mathbb{C}^-$ was obtained by Stieltjes in 1889 \cite{Stieltjes}.
 
The series (\ref{Gudermann-2}) converges for all $z\in\mathbb{C}^-$
and $\mu$ can be bounded by
\begin{equation}
|\mu(z)| \leq \frac{1}{12}\frac{1}{\cos^2(\varphi/2)}\frac{1}{|z|},
\label{bound-for-mu}
\end{equation}
with $z=|z|e^{i\varphi}$, $z\in\mathbb{C}^-$ (see \cite{Remmert} or
\cite{Remmert-on-Wielandt}). On the real line, very precise upper and
lower approximants of the $\Gamma$-function can be obtained from
(\ref{Gudermann-1}) and (\ref{Gudermann-2}), valid also for small values
of the argument (see \cite{HRobbins} and \cite{Mansour}).

Notice that, by (\ref{bound-for-mu}), for $|z|\simeq 1$, the
contribution of the factor $e^{\mu(z)}$ to (\ref{Gudermann-1}) is
lower than ci.\ $9\%$ for $\varphi\simeq 0$ and lower than ci.\ $18\%$
$\varphi\simeq \pi/2$. Hence, for the sake of precision, it is
relevant in that range to consider corrections to Stirling's
approximation (\ref{Gamma-large-l}).

With (\ref{Gudermann-1}), and since $\Gamma(z+1)=z\Gamma(z)$, we can also write
\begin{equation}
\Gamma(z+1) = \sqrt{2\pi}\, e^{-z}\, z^{z+1/2}\,e^{\mu(z)} .
\label{Gudermann-1b}
\end{equation}
Hence, $e^{\mu(z)}$ acts as a correcting factor for both
(\ref{Gamma-large-l}) and (\ref{Gamma-large-l-A}).  Beyond the
Gudermann series (\ref{Gudermann-2}), the function $\mu$ can be
represented in many other forms. One of the most useful of them is the
so-called Stirling series:
\begin{equation}
\mu(z) = \sum_{n=1}^{\infty} \frac{B_{2n}}{(2n-1)2n}\frac{1}{z^{2n-1}},
\label{Stirling-for-mu}
\end{equation}
where $B_k$ is the $k$-th Bernoulli number. One has to say that the
series on the r.h.s.\ of (\ref{Stirling-for-mu}) is an asymptotic
series in $\mathbb{C}^-$, but is not convergent! See again
\cite{Remmert}. Therefore, it is not to be seen as a Laurent expansion
for $\mu$ around $z=0$. In fact, $z=0$ is not an isolated singularity
of $\mu$, but a branch point, as we see from
(\ref{Gudermann-2}). Since (\ref{Stirling-for-mu}) is an asymptotic
series we can get good approximations for $\mu$ in $\mathbb{C}^-$ by
truncating it at some finite value of the index $n$ and taking $1/|z|$
small enough. The first terms of (\ref{Stirling-for-mu}) are
\begin{equation}
\mu(z) = \frac{1}{12}\frac{1}{z} - \frac{1}{360}\frac{1}{z^3} + \frac{1}{1260}\frac{1}{z^5}
+ \cdots .
\label{Stirling-para-mu}
\end{equation}
For the correcting factor $e^{\mu(z)}$, this gives
\begin{equation}
e^{\mu(z)} = 1 + \frac{1}{12}\frac{1}{z} + \frac{1}{288}\frac{1}{z^2} 
            - \frac{139}{51840}\frac{1}{z^3} + \cdots .
\end{equation}
Therefore, we may write
\begin{subequations}
\label{Stirling}
\begin{eqnarray}
\Gamma(z) & = & \sqrt{2\pi}\, e^{-z}\, z^{z-1/2}\, \left[1 +  \frac{1}{12}\frac{1}{z} + \frac{1}{288}\frac{1}{z^2} 
            - \frac{139}{51840}\frac{1}{z^3} + \cdots\right],
\label{Stirling-I}
\\
\Gamma(z+1) & = & \sqrt{2\pi}\, e^{-z}\, z^{z+1/2}\, \left[1 + \frac{1}{12}\frac{1}{z} + \frac{1}{288}\frac{1}{z^2} 
            - \frac{139}{51840}\frac{1}{z^3} + \cdots\right] .
\label{Stirling-II}
\end{eqnarray}
\end{subequations}

These approximations are also known as Stirling's series for the gamma
function. They are asymptotic (but not convergent!)  expansions for
$\Gamma$, valid for $z\in\mathbb{C}^-$ and $1/|z|$ ``small''. Although
(\ref{Stirling}) are asymptotic but not convergent approximations for
$\Gamma$, they can in some practical situations be more useful than the
representation (\ref{Gudermann-1}) or (\ref{Gudermann-1b}) with the
convergent Gudermann expansion (\ref{Gudermann-2}).

\section{Stirling's series. A formal derivation }

The Stirling's series (\ref{Stirling}) leads to more accurate
estimates of the phase-shifts (\ref{sigma-gamma}) than Stirling's
approximation (\ref{Gamma-large-l}), even when the value of $l$ is not
``large''.
There are many proofs of Eqs.~(\ref{Stirling-for-mu}) or
(\ref{Stirling}) in the real or in the complex domain (see, e.g.,
Ref.\ \cite{Remmert, Hardy-DivergentSeries}), but they are all rather
involved. We will present now a simple derivation of
Eq.~(\ref{Stirling-II}) (see also \cite{Hochstadt1}).

Consider the curve $C$ in the complex $t$-plane parametrized by 
$s\in(-\infty, \; \infty)$ and satisfying the parabolic functional mapping
\cite{Dingle73, HP97, Hochstadt1},
$$
t(s) - z\ln\big(t(s)\big) - A(z) =\frac{s^{2}}{2} ,
$$
with $A(z) = z-z\ln{z}$ and with $t(0)=z$. Defining
$
\omega(t) = t - z\ln(t) ,
$
we can write
$$
\omega\big(t(s)\big) =  A(z) + \frac{s^{2}}{2} .
$$
One can choose $t(s)$ satisfying $t(s)\to 0$ as $t(s)\simeq
e^{-\frac{s^2}{2z}}$ for $s\to-\infty$ and $t(s)\simeq \frac{s^{2}}{2}
+ i\mbox{Im}(z)\ln\left( \frac{s^2}{2}\right)$ for $s\to+\infty$. By a
continuous deformation of the $t$-integration curve in
(\ref{EulerIntegral}) from the positive real axis to $C$ and by
carefully extending the integration to infinity, one can write, by
Cauchy's theorem,
$$
\Gamma(1+z)=\int_C dt\, t^{z}\, e^{-t} = 
e^{-z+z\ln{z}}\, \int_{-\infty}^{\infty}\, e^{-s^{2}/2}\, \frac{d t(s)}{ds}\, ds.
$$

The derivative $dt/ds$ can be written as a power series in $s^2$
using the mapping equation,
\begin{equation}
\frac{dt}{ds}=\sum_{k=0}^{\infty}\, a_{2k}\, s^{2k}\label{dt-ds_exp}.
\end{equation}
It is a simple matter to evaluate the coefficients $a_k$, by repeated
differentiation of the mapping equation. We calculate below the first
two terms, $a_0$, and $a_2$ (higher order terms can be derived similarly).
Calling $\omega\big(t(s)\big)\equiv v(s)$, we write
\begin{eqnarray*}
  \frac{dv(s)}{ds}&=&\left(1-\frac{z}{t}\right)\, \frac{dt}{ds}=s,
  \\
  \frac{d^2v(s)}{ds^2}&=&
  \left(1-\frac{z}{t}\right)\, 
  \frac{d^{2}t}{ds^2}+\frac{z}{t^2}\, \left(\frac{dt}{ds}\right)^2
  = 1,
  \\
  \frac{d^3v(s)}{ds^3}&=&
  \left( 1-\frac{z}{t} \right)\, \frac{d^{3}t}{ds^3}
   +\frac{2z}{t^2}\, \frac{d^{2}t}{ds^2}\, \frac{dt}{ds}-
    \frac{2z}{t^3}\, \left(\frac{dt}{ds}\right)^3=0,
  \\
  \frac{d^4v(s)}{ds^4}  &=&
  \left( 1-\frac{z}{t} \right)\, \frac{d^{4}t}{ds^4}
  + \frac{4z}{t^2}\, \frac{d^{3}t}{ds^3}\,\frac{dt}{ds} 
  +\frac{3z}{t^2}\, \left(\frac{d^{2}t}{ds^2}\right)^2
  \\ & &
  -\frac{12 z}{t^3}\, \frac{d^{2}t}{ds^2}\left(\frac{dt}{ds}\right)^2
  +\frac{6z}{t^4}\, \left(\frac{dt}{ds}\right)^4 = 0.
\end{eqnarray*}
We now evaluate these equations at the extremum point, defined by the condition
\begin{equation}
\left[ \frac{d\omega(t)}{dt} \right]_{t_0} =0, 
\end{equation}
which is $ t_0=z$ and $s =0$. The above four equations lead to the results
\begin{equation}
\left[ \frac{dt}{ds}\right]_{t=t_0}=\sqrt{z},
\; 
\left[ \frac{d^3t}{ds^3}\right]_{t=t_0} = \frac{1}{6\, \sqrt{z}}
\;{\rm and}\;
\left[\frac{d^5}{ds^5}\right]_{t=t_0} = \frac{1}{36\,z^{3/2}}.
\end{equation}
Thus the coefficients $a_{2k}$ above are,
\begin{equation}
a_0 = \left[ \frac{dt}{ds}\right]_{t=t_0} = \sqrt{z},
\end{equation}
\begin{equation}
a_2 = \frac{1}{2!}\, \left[ \frac{d^3t}{ds^2}\right]_{t=t_0} = \frac{1}{12\, \sqrt{z}},
\end{equation}
\begin{equation}
a_4 = \frac{1}{4!}\, \left[\frac{d^{5}t}{ds^5}\right]_{t=t_0} = \frac{1}{4!}\ \frac{1}{36\,z^{3/2}}.
\end{equation}
Using the Gaussian integral formula,
\begin{equation}
\int_{-\infty}^{\infty} ds\, e^{-s^2/2}\, s^{n}\,ds =\sqrt{2\pi}\,  (n-1)!!,
\end{equation}
for $n$ even, the full integral in the $\Gamma$ function can be
written down in the form of the Stirling series above, namely,
\begin{equation}
\Gamma(z+1) = \sqrt{2\pi}\, e^{-z}\, z^{z}\, 
\sum_{k=0}^{\infty}(2k-1)!!\, a_{2k}.
\end{equation}
Thus, evaluating the coefficients $a_{2k}$ and inserting above, we get
\begin{equation}
\Gamma(z+1) 
=
\sqrt{2\pi}\, \, e^{-z}\, z^{z+1/2}\, 
\left[1+\frac{1}{12\, z}+\frac{1}{288\, z^2}+\cdots\right] .
\label{ch3_series}
\end{equation}

The use of the quadratic mapping analysis of integrals and the
generation of appropriate asymptotic series in general
was demonstrated in, e.g., \cite{HP97} for the case of the Gamow
integral employed in nuclear astrophysics. This integral has the general form,
\begin{equation}
I(a)= \int_{0}^{\infty}dx S(x)\exp{\left(-x -\frac{a}{x^{1/2}}\right)},
\end{equation}
where the function $S(x)$ is usually a slowly varying function of $x$
and can be written as a sum $S(x) =\sum_{k=0}^{\infty}S_{k}x^{k}$, which
would then result in a series representation of the Gamow integral
similar to what was done for the $\Gamma$ function above.\\

We now use the above results to evaluate the Coulomb phase-shifts
(\ref{sigma-gamma}).

\section{An exact expression for the phase shift and first approximations}

From (\ref{sigma-gamma}) and (\ref{Gudermann-1}), and using the
identity
\begin{equation}
\ln\left( \frac{x+iy}{x-iy}\right) 
=
2i \tan^{-1}\left( \frac{y}{x} \right) ,
\label{tan-1eln}
\end{equation}
valid for $x, \; y\in\mathbb{R}$ with $x>0$, we get after some
elementary computations the following \underline{exact} expression for
the phase shift:
\begin{equation}
\sigma_{l}  = \sigma_{l}^{(0)} + M_{l,\,\eta} ,
\label{FormulaExataParaSigmal-A}
\end{equation}
where we define
\begin{equation}
\sigma_{l}^{(0)}  \equiv
\left( l+ \frac{1}{2}\right)\tan^{-1}\left( \frac{\eta}{l+1} \right)
+ \eta\left( \ln\left(\sqrt{(l+1)^2 + \eta^2} \right) -1 \right) ,
\label{FormulaExataParaSigmal-B}
\end{equation}
and $M_{l,\,\eta}\equiv \frac{1}{2i}\big( \mu(1+l+i\eta) -
\mu(1+l-i\eta)\big)$. Below, we will use the series expansion
(\ref{Gudermann-2}) to find closed expressions for $M_{l,\,\eta}$, for
$\sigma_l$ and, in particular, for $\sigma_0$. Before we proceed let
us make some comments about some useful approximation we can obtain
from
(\ref{FormulaExataParaSigmal-A})--(\ref{FormulaExataParaSigmal-B}).

Using
(\ref{bound-for-mu}) with $|z|=\sqrt{(l+1)^2 + \eta^2}$ and
$\varphi=\tan^{-1}\left( \frac{\eta}{l+1} \right)$, one finds the
bound
$$
|M_{l,\,\eta}| 
\leq 
\frac{1}{6\Big(l+1+\sqrt{(l+1)^2 + \eta^2}\Big)}.
$$
Hence, for $l\gg 1$ or $|\eta|\gg 1$ the contribution of
$M_{l,\,\eta}$ to (\ref{FormulaExataParaSigmal-A}) can be neglected and
we can restrict as a first approximation to $\sigma_l^{(0)}$.

For $l=0$, for instance, one gets from
(\ref{FormulaExataParaSigmal-B}) the approximation
\begin{equation}
\sigma_{0}
\simeq
\sigma_{0}^{(0)}
=
\frac{1}{2}\tan^{-1}\left( \eta \right)
+ \eta\left( \ln\left(\sqrt{ 1 + \eta^2} \right) -1\right)
\label{sigmal-for-l-eq-0}
\end{equation}
with an error bounded by
$\frac{1}{6}\Big(1+\sqrt{1 + \eta^2}\Big)^{-1}$.
For $|\eta|\gg 1$, Eq.~(\ref{sigmal-for-l-eq-0}) gives the approximation
\begin{equation}
\sigma_{0}
\simeq
\frac{\pi}{4}
+\eta\big( \ln(\eta) -1 \big).
\label{sigmal-for-l-eq-0-approx}
\end{equation}

For  $\frac{|\eta|}{l+1}\ll 1$ and $l\gg1$ we get from 
(\ref{FormulaExataParaSigmal-B}) the approximation
\begin{equation}
\sigma_{l} 
\simeq
\eta \ln(l+1).
\label{glauber}
\end{equation}

\section{Closed expression for the phase shift }

Now we will try to find closed expressions for $\sigma_l$ and
$\sigma_0$.  According to (\ref{sigma-gamma}), using the fact that
$\Gamma(z+1)=z\Gamma(z)$ for all $z\in\mathbb{C}^-$, we have
$$
e^{2i\sigma_l}
=
\frac{\Gamma(1+l+i\eta)}{\Gamma(1+l-i\eta)}
=
\frac{(l+i\eta) \cdots (1+i\eta)}{(l-i\eta) \cdots (1-i\eta)}
\frac{\Gamma(1+i\eta)}{\Gamma(1-i\eta)}
=
\frac{(l+i\eta) \cdots (1+i\eta)}{(l-i\eta) \cdots (1-i\eta)}e^{2i\sigma_0}.
$$
Therefore $\sigma_l = \sigma_0 + \frac{1}{2i}\sum_{m=1}^l\ln\left(
  \frac{m+i\eta}{m-i\eta} \right)$ and using (\ref{tan-1eln}), one has
\begin{equation}
\sigma_l
=
\sigma_0 + \sum_{m=1}^l\tan^{-1}\left( \frac{\eta}{m} \right).
\label{somafinitaparasigma-l}
\end{equation}
Eq.\ (\ref{somafinitaparasigma-l}) is a remarkable expression, since
it shows that $\sigma_l$ differs from $\sigma_0$ by a finite sum.
Let us now find a more explicit expression for $\sigma_0$. According
to (\ref{FormulaExataParaSigmal-A})--(\ref{FormulaExataParaSigmal-B}),
\begin{equation}
\sigma_{0}  = 
\frac{1}{2}\tan^{-1}\left( \eta \right)
+ \eta\left(\ln\left( \sqrt{1 + \eta^2 } \right) -1 \right)
+ M_{0,\,\eta} ,
\label{FormulaExataParaSigma-0}
\end{equation}

Now we analyze $M_{0,\,\eta}\equiv \frac{1}{2i}\big( \mu(1+i\eta)
- \mu(1-i\eta)\big)$ more closely. According to (\ref{Gudermann-2})
(with the change of summation variable $m\to m-1$), we have,
\begin{equation}
M_{0,\,\eta} =
\frac{1}{2i}\sum_{m=1}^\infty
\Bigg[
\left( m + i\eta  +\frac{1}{2}\right) \ln\left(1+\frac{1}{m + i\eta }\right)
-
\left( m - i\eta +\frac{1}{2}\right) \ln\left(1+\frac{1}{m - i\eta }\right)
\Bigg].
\label{AUX-0}
\end{equation}
After simple rearrangements and using  we can write (\ref{AUX-0}) as
\begin{equation}
M_{0,\,\eta} =
\frac{1}{2i}\sum_{m=1}^\infty
\left[
\left( m +\frac{1}{2}\right) 
\left( 
  \ln\left( \frac{m+1+i\eta}{m+1-i\eta}\right) 
  - \ln\left(\frac{m+i\eta}{m-i\eta}\right)
\right)
+i\eta\Bigg( 
        \ln\left((m+1)^2 + \eta^2 \right) -  \ln\left(m^2 + \eta^2 \right) 
      \Bigg)
\right].
\label{AUX-1}
\end{equation}
Using (\ref{tan-1eln}) and defining
$$
A_m \equiv 
\frac{1}{2i} \ln\left(\frac{m+i\eta}{m-i\eta}\right)
= \tan^{-1}\left( \frac{\eta}{m} \right)
\qquad \mbox{ and } \qquad
B_m \equiv \frac{\eta}{2}\ln\left(m^2 + \eta^2 \right) 
$$
we can write (\ref{AUX-1}) as $M_{0,\,\eta} = \lim_{N\to\infty}M_{0,\,\eta}^N$, where
\begin{equation}
M_{0,\,\eta}^N =
\sum_{m=1}^N
\left[
\left( m +\frac{1}{2}\right) (A_{m+1} - A_m)
+B_{m+1} - B_m
\right].
\label{AUX-3}
\end{equation}
Now,
\begin{equation}
\sum_{m=1}^N
\left( B_{m+1} - B_m\right)
=
B_{N+1} - B_{1}
\end{equation}
and
\begin{eqnarray}
\sum_{m=1}^N
\left( m +\frac{1}{2}\right) (A_{m+1} - A_m)
& = &
\sum_{m=1}^N
\left[
\left( (m + 1) + \frac{1}{2}\right) A_{m+1}
-
\left( m + \frac{1}{2}\right) A_{m}
\right]
-
\sum_{m=1}^N  A_{m+1}
\nonumber\\
& = &
\left( N + 1 + \frac{1}{2}\right) A_{N+1}
-
\left(  1 + \frac{1}{2}\right) A_{1}
-
\sum_{m=1}^N  A_{m+1} .
\end{eqnarray}
Hence, collecting the results, we have
\begin{eqnarray}
M_{0,\,\eta} 
& = &
\lim_{N\to\infty}
\left[ 
\left(N+\frac{3}{2}\right) \tan^{-1}\left( \frac{\eta}{N+1} \right)
+ \frac{\eta}{2} \ln\Big( (N+1)^2 + \eta^2 \Big)
-\sum_{m=1}^N \tan^{-1}\left( \frac{\eta}{m+1} \right)
\right]
\nonumber \\
& &
-\frac{3}{2} \tan^{-1}\left( \eta \right)
-\frac{\eta}{2} \ln\Big( 1 + \eta^2 \Big)
\nonumber \\
& = &
 \eta -\frac{3}{2} \tan^{-1}\left( \eta \right)
-\frac{\eta}{2} \ln\Big( 1 + \eta^2 \Big)
+\lim_{N\to\infty}
\left[ 
\eta\ln(N+1) - \sum_{m=1}^N \tan^{-1}\left( \frac{\eta}{m+1} \right)
\right].
\label{AUX-4}
\end{eqnarray}
Now, adding and subtracting $\eta\sum_{m=0}^N \frac{1}{m+1}$ to the
terms in brackets, whose limit is being taken in (\ref{AUX-4}), we get
$$
\eta\ln(N+1) - \sum_{m=1}^N \tan^{-1}\left( \frac{\eta}{m+1} \right)
= 
\eta\left[\ln(N+1) - \sum_{m=0}^N  \frac{1}{m+1} \right]
 - 
\sum_{m=0}^N \left[\tan^{-1}\left( \frac{\eta}{m+1}\right) - \frac{\eta}{m+1}\right]
+ \tan^{-1}\left( \eta \right) .
$$
With this, (\ref{AUX-4}) becomes
\begin{equation}
M_{0,\,\eta} 
 = 
 \eta(1-\gamma) -\frac{1}{2} \tan^{-1}\left( \eta \right)
-\frac{\eta}{2} \ln\Big( 1 + \eta^2 \Big)
- \sum_{m=0}^\infty \left[\tan^{-1}\left( \frac{\eta}{m+1}\right) -  \frac{\eta}{m+1}\right] ,
\label{AUX-5}
\end{equation}
where $\gamma\equiv \lim_{N\to\infty} \left[ \sum_{m=0}^N
  \frac{1}{m+1} - \ln(N+1) \right]$ is Euler's constant $\gamma\simeq 0.5772156649\ldots$.

Inserting (\ref{AUX-5}) into (\ref{FormulaExataParaSigma-0}) we finally
get, after some trivial cancelations,
\begin{equation}
\sigma_0 = 
-\gamma\eta - \sum_{m=0}^\infty \left[\tan^{-1}\left(
    \frac{\eta}{m+1}\right) - \frac{\eta}{m+1}\right] ,
\label{FINAL-sigma-0}
\end{equation}

By recalling the Taylor expansion of $\tan^{-1}$ about $0$,
\begin{equation}
\tan^{-1}(x)=\sum_{k=0}^{\infty}\frac{(-1)^k}{2k+1}x^{2k+1} ,
\label{Tylor-for-cotan}
\end{equation}
we can write (\ref{FINAL-sigma-0}) for $|\eta|<1$ as
$$
\sigma_0 = 
-\gamma\eta 
- \sum_{m=0}^\infty \sum_{k=1}^{\infty}
\frac{(-1)^k}{2k+1} \left(  \frac{\eta}{m+1}\right)^{2k+1}
=
-\gamma\eta - \sum_{k=1}^{\infty}\frac{(-1)^k\zeta(2k+1)}{2k+1} \eta^{2k+1},
$$
where $\zeta$ is Riemann's zeta function.

\begin{figure}[ptb]
\begin{center}
\includegraphics*[height=6cm]{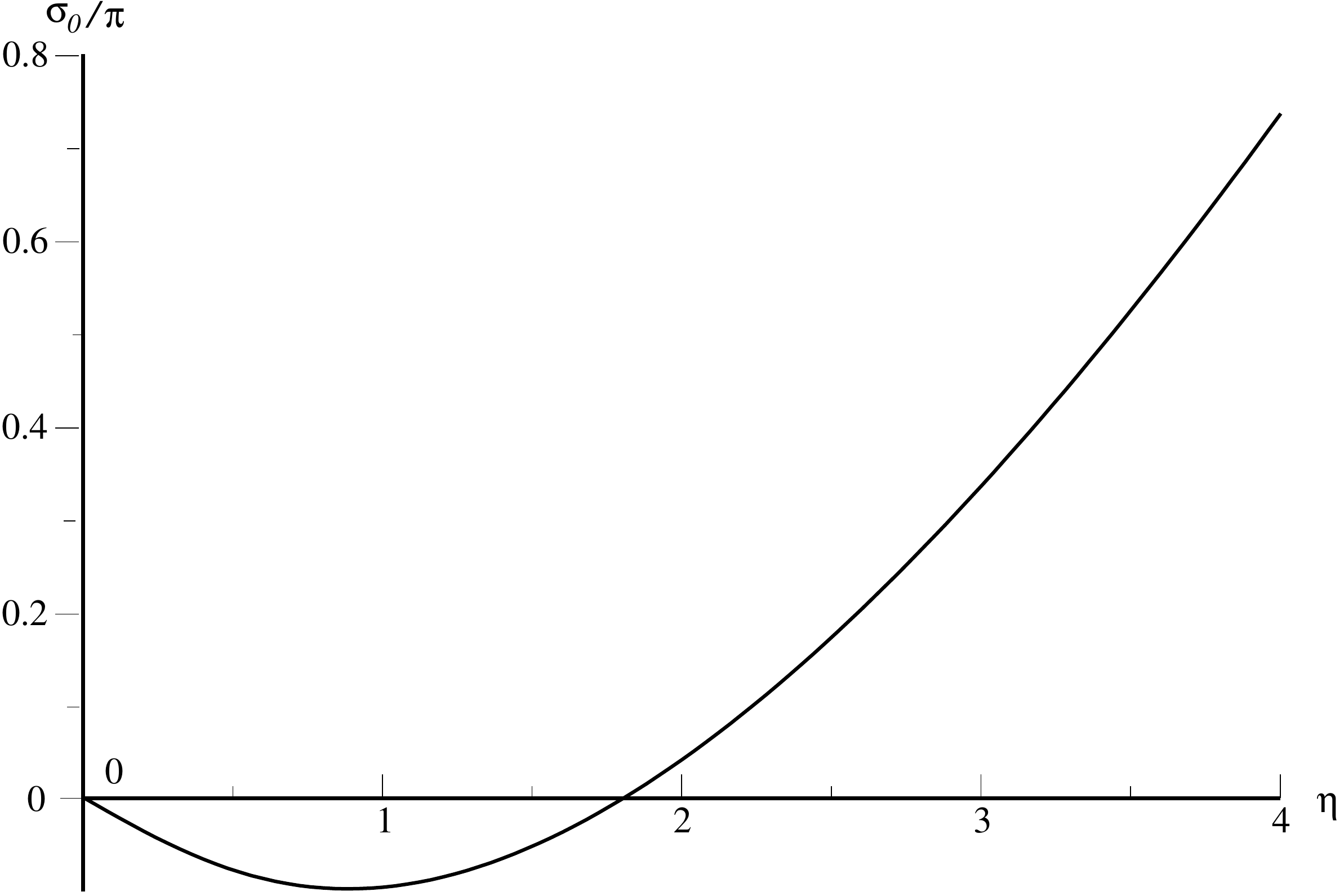}
\end{center}
\caption{The graph of $\sigma_0/\pi$ according to the exact formula
  (\ref{FINAL-sigma-0}) for $0\leq \eta \leq 4$. Notice that
  $\sigma_0$ vanishes at $\eta \simeq 1.810$.}
\label{sigma0-exata}
\end{figure}

In Figure \ref{sigma0-exata} we plot $\sigma_0$ for $0\leq \eta \leq
4$. As we discuss below, for larger values $\sigma_0$ follows very closely the behavior
dictated by the asymptotic approximations 
(\ref{sigmal-for-l-eq-0}) or (\ref{sigmal-for-l-eq-0-approx}). 
In Figure \ref{sigma0-exata} we see that $\sigma_0$ vanishes at
$\eta=0$ and $\eta \simeq 1.810$. This latter value of $\eta$ is coincidentally quite close to the value of $\sqrt{2}$, which was obtained in \cite{CDH01} in connection with the scattering of identical charged particles (Fermions or Bosons). The Mott cross section, at this
critical value of $\eta = \sqrt{2}$, was predicted to be isotropic over a broad range of angles around $90^{o}$.  

\section{The phase shift and Stirling's series}

The above asymptotic formulae (\ref{FormulaExataParaSigmal-B}),
(\ref{sigmal-for-l-eq-0}), (\ref{sigmal-for-l-eq-0-approx}) and
(\ref{glauber}) will be discussed further below
within the low-energy, large-$\eta$, large-$l$ case of the WKB
approximation, and the high-energy, small-$\eta$, large-$l$ case of
the eikonal approximation. In this section we use Stirling series 
for the gamma function (\ref{Stirling}) to further improve those
approximations. 

\begin{figure}[ptb]
\begin{center}
\includegraphics*[height=6cm]{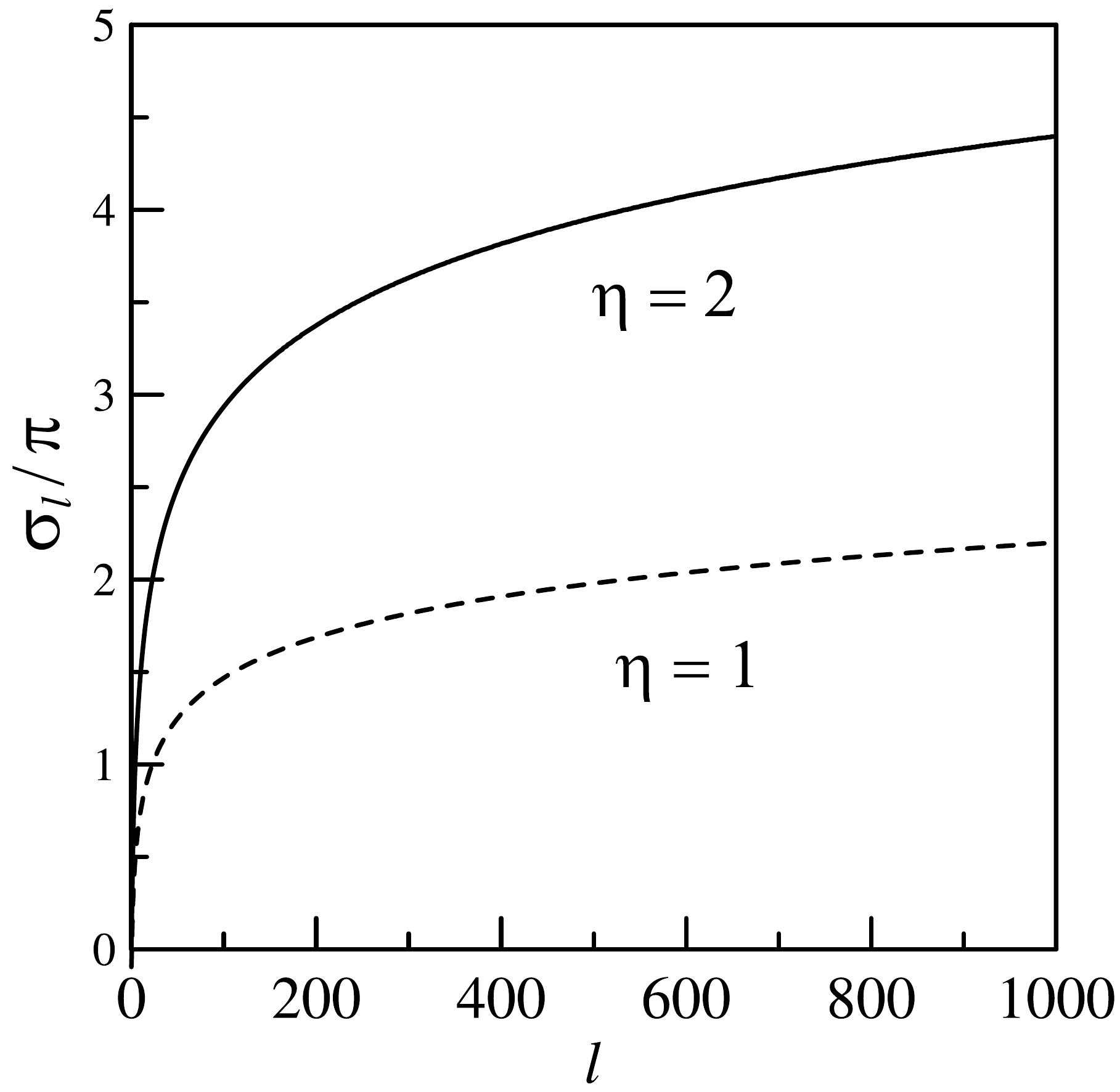}
\end{center}
\caption{Coulomb phase-shifts as functions of $l$. Results normalized with
respect to $\pi$ are given for two values of the Sommerfeld parameter.}
\label{coul_l}
\end{figure}

A first order correction to Eq.~(\ref{FormulaExataParaSigmal-B}) can
be easily evaluated using Stirling's series
(\ref{Stirling-para-mu})--(\ref{Stirling}). For this purpose, we
rewrite Eq.~(\ref{sigma-gamma}) in the form
\begin{equation}
e^{2i\sigma_l}= e^{2i\sigma_l^{(0)}} \, F(z),
\end{equation}
with $z=1+l+i\eta$ and
$\sigma_l^{(0)}$ given in (\ref{FormulaExataParaSigmal-B}), 
where
\begin{equation}
F(z)= e^{\mu(z)-\mu(z^*)}
=
\exp\left( 
         \frac{1}{12}\left( \frac{1}{z} - \frac{1}{z^*} \right) 
         -\frac{1}{360}\left( \frac{1}{(z)^3} - \frac{1}{(z^*)^3} \right) +\cdots\right) .
\label{F(z)}
\end{equation}
where the $*$ symbol refers to complex conjugation.
The first order approximation for $F(z)$ is
\begin{equation}
F^{(1)}(z)\equiv 
\exp\left( \frac{1}{12}\left( \frac{1}{z} - \frac{1}{z^*} \right) \right) 
=: e^{2i\, \Delta\sigma_l^{(0)}}.
\label{F1(z)}
\end{equation}
Evaluating the above expression, we find
\begin{equation}
\Delta\sigma_l^{(0)}
= - \frac{\eta} {12\,\big((l+1)^2+\eta^2\big)}   .
\label{Delta_sigma0}
\end{equation}
Therefore, the first order approximation to the Coulomb phase-shifts is
\begin{eqnarray}
\sigma_l^{(1)} & = & \sigma_l^{(0)}+\Delta\sigma_l^{(0)}
\nonumber \\
& = &
\left( l+ \frac{1}{2}\right)\tan^{-1}\left( \frac{\eta}{l+1} \right)
+ \eta\bigg( \ln\Big(\sqrt{(l+1)^2  + \eta^2} \Big) -1\bigg)
- \frac{\eta} {12\,\big((l+1)^2+\eta^2\big)} .
\label{Sigma1_l}
\end{eqnarray}
Figure \ref{coul_l} shows Coulomb phase-shifts versus $l$. 

\begin{figure}[ptb]
\begin{center}
\includegraphics*[height=6.0cm]{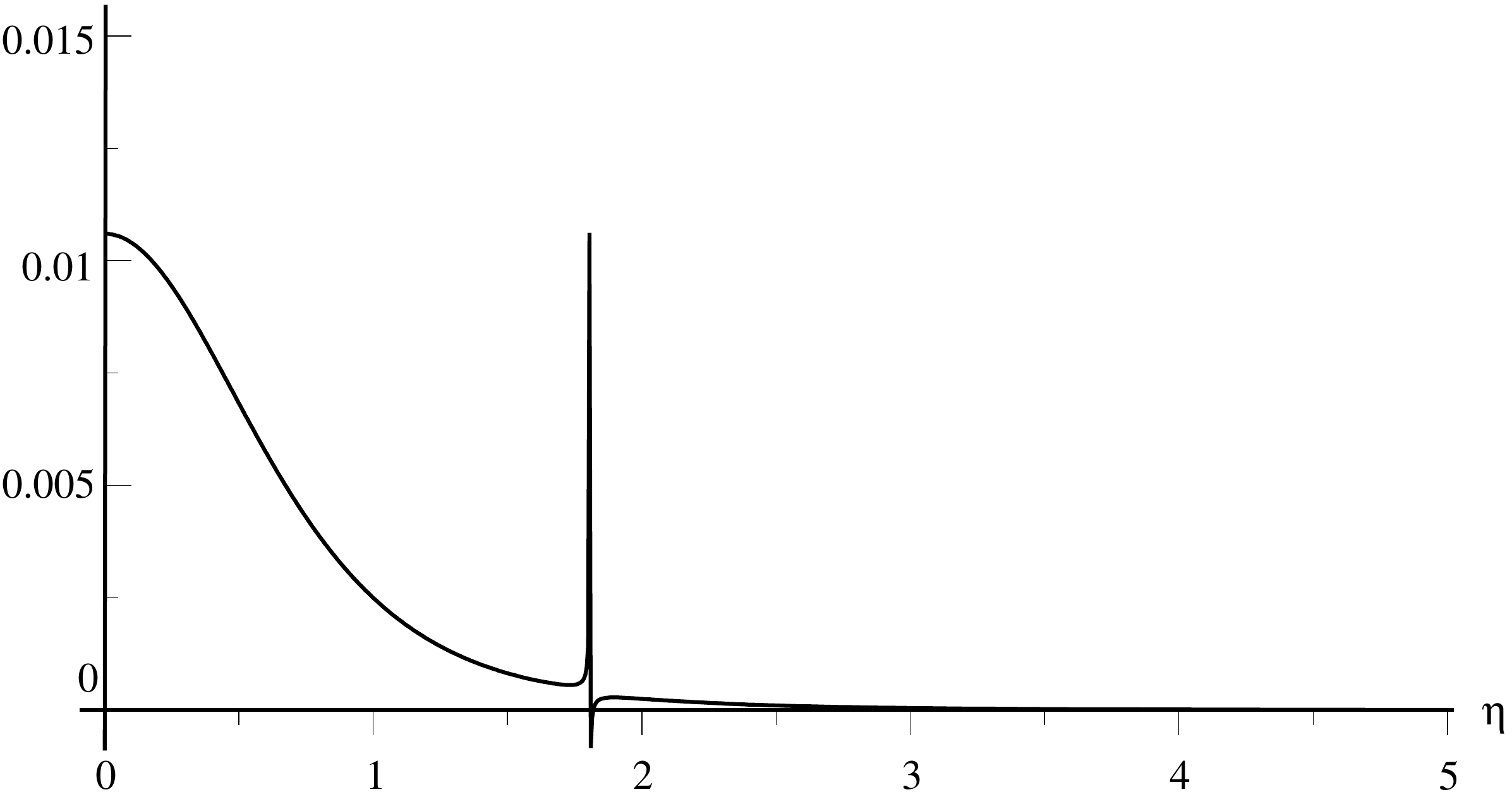}
\end{center}
\caption{The relative error $\frac{\sigma_0^{(1)} - \sigma_0}{\sigma_0}$
  for $0\leq \eta\leq 5$.
The sharp peak around $\eta\simeq 1.81$ is due the the vanishing of
$\sigma_0$ at that point. }
\label{sigma1_0-MENOS-sigma-0-RELATIVO}
\end{figure}

From (\ref{Sigma1_l}) we get, in particular, the first order correction to 
(\ref{sigmal-for-l-eq-0}) due to Stirling's series:
\begin{eqnarray}
\sigma_0^{(1)} & = & \sigma_0^{(0)}+\Delta\sigma_0^{(0)}
\nonumber \\
& = &
\frac{1}{2}\tan^{-1}\left( \eta \right)
+ \eta\bigg( \ln\left(\sqrt{ 1 + \eta^2 }\right) -1\bigg)
- \frac{\eta} {12\,\left(1+\eta^2\right)} .
\label{Sigma1_0}
\end{eqnarray}
It is interesting to compare the approximation (\ref{Sigma1_0}) to our
exact expression for $\sigma_0$ given in
(\ref{Formula-exata-para-sigma-0-I}).  Figure
\ref{sigma1_0-MENOS-sigma-0-RELATIVO} we plot the relative error
$\frac{\sigma_0^{(1)}-\sigma_0}{\sigma_0}$ for values of the Sommerfeld
parameter between $0$ and $5$. It shows that $\sigma_0^{(1)}$ is an
excellent approximation for $\sigma_0$, with relative errors below
$1\%$, even for ``small'' values of $\eta$, except, perhaps, near
$\eta\simeq 1.81$, where $\sigma_0$ vanishes.

\section{Discussion of the results for Stirling's series}

\begin{figure}[ptb]
\begin{center}
\includegraphics*[height=8cm]{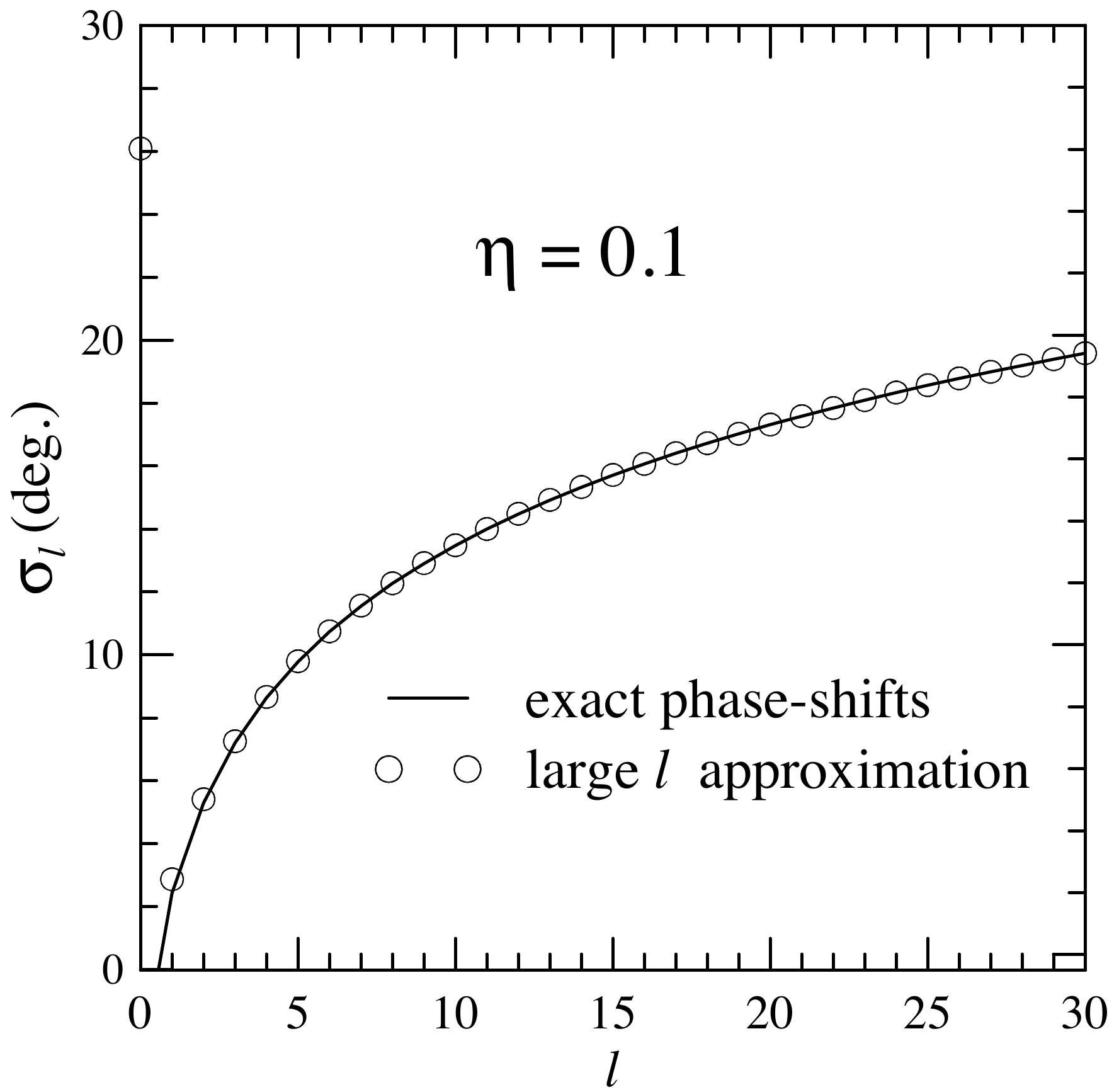}
\end{center}
\caption{The approximate Coulomb phase-shifts $\sigma_l^{(0)}$ and
  $\sigma_l^{(1)}$ for $\eta=0.1$, as functions of $l$. For details
  see the text.}
\label{ch3_large-l}
\end{figure}

The lowest orders approximations (in $1/z$) of the previous section
are supposed to work for $| z | \gg 1$, which means $l \gg 1$ and/or
$\eta\gg 1$. They are, in fact, very accurate, even when these
conditions are not well satisfied. A first illustration of this fact
is presented in Figure \ref{ch3_large-l}, where we compare the exact
phase-shifts (solid line) with the lowest order approximation
$\sigma_l^{(0)}$ (open circles) for a small $\eta$, as functions of
$l$. They are very close. The only exception is the case of $l=0$,
where they are quite different.

For larger values of $\eta$, the agreement is much better. This can be
seen in Table \ref{tab-sigcou}, where we compare the two
approximations of the previous section with the exact Coulomb
phase-shifts.  Note that the $\sigma_l^{(1)}$ is very accurate even
for $l=0$ and $\eta=0.1$.

\begin{table}
\centering
\begin{tabular} [c] {ccccc}
\hline
& & & &\\ 
$\quad\eta\quad$  & $\quad  l\quad$  & $\quad\sigma_l^{(0)}/\pi$ $\quad$ & $
\quad\sigma_l^{(1)}/\pi\quad$ & $\sigma_l^{\rm exact}/\pi$ \\
& & & &\\ 
 \hline
& & & &\\ 
  0.1 &  0 & -0.01581  &  -0.01844  &  -0.01825 \\
         &  1 &  0.01413  &  0.01346   &  0.01348  \\
         &  2 &  0.02967  &  0.02930   &  0.02938  \\
& & & &\\ 
\hline 
& & & &\\  
 1.0 & 0 & -0.08299 & -0.09625 & -0.09602 \\
     & 1    & 0.1592 & 0.1539  & 0.1540 \\
     & 2    &  0.3042 & 0.3015  & 0.3016 \\
\hline
\end{tabular}
\caption{Coulomb phase-shifts as functions of
  $l$, for two values of the Sommerfeld parameter.}
\label{tab-sigcou}
\end{table}
%

\begin{figure}[ptb]
\begin{center}
\includegraphics*[height=8 cm]{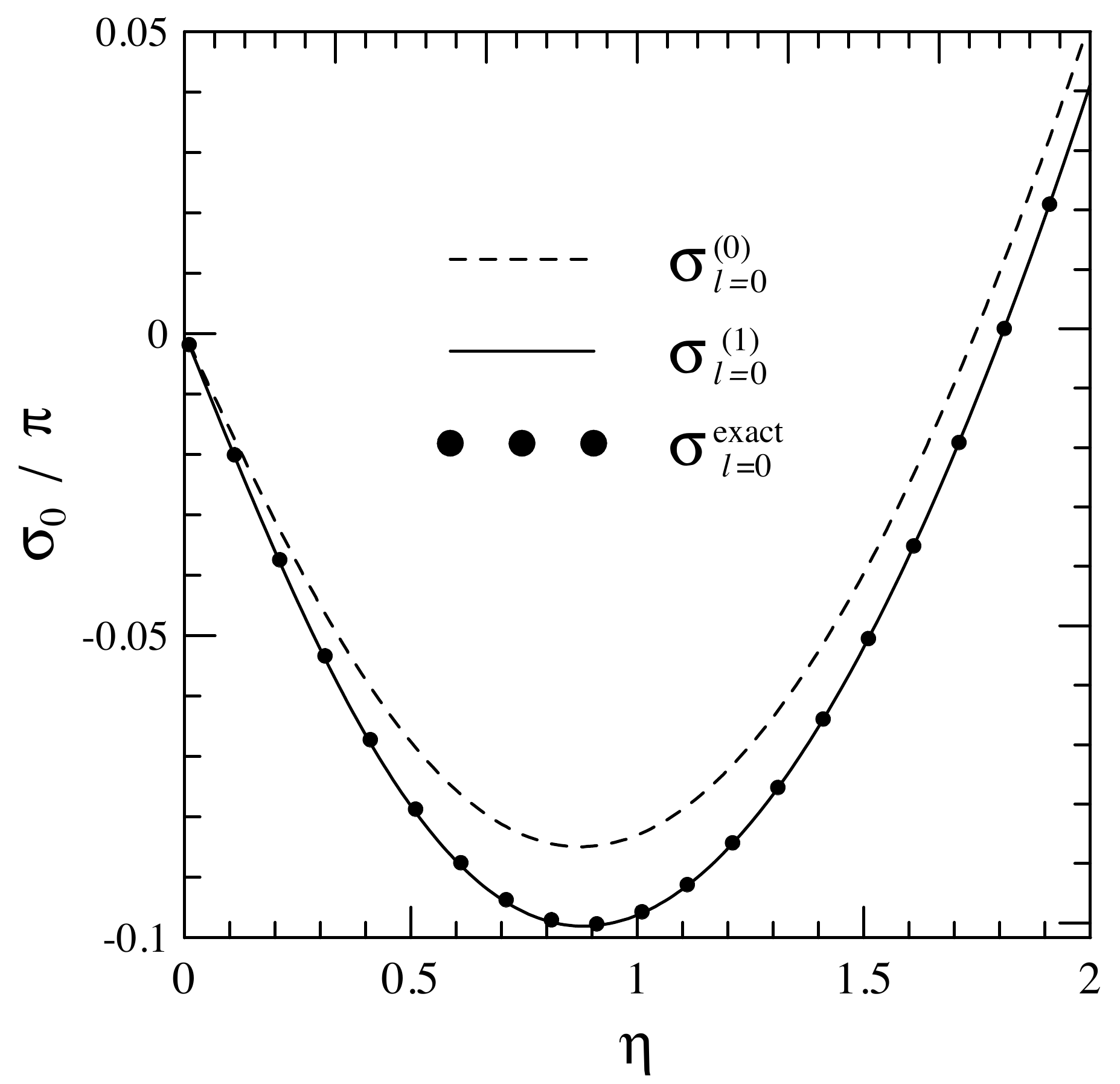}
\end{center}
\caption{Coulomb s-wave phase-shifts as a function of the Sommerfeld
  parameter. Exact values are compared with the approximations
  discussed in the text. }
\label{s-wave-approx}
\end{figure}

This point can be seen more clearly in Figure \ref{s-wave-approx},
where the exact s-wave phase-shifts are compared with the
approximations $\sigma_{l=0}^{(0)}$ and $\sigma_{l=0}^{(1)}$. The comparison
indicates that the usual large-$\eta$ approximation, $\sigma_{l=0}^{(0)}$, is
very poor. The situation is completely different with the improved
approximation, $\sigma_{l=0}^{(1)}$, which includes the first order
correction of Stirling's series. In this case, one gets accurate
results for any value of the Sommerfeld parameter, $\eta$.

\section{Comparison with other approximations}

We turn now to well known approximations used to calculate the phase
shifts, motivated by the physical conditions. The first such
approximation is the WKB one invoked to consider scattering under
semi-classical conditions of short local wave lengths. In this
approximation, one finds for the phase shift the following expression
\cite{Brink85},
\begin{equation}
\sigma_{l} = 
\lim_{r \rightarrow \infty}
\left[\int_{r_{l}}^{r}dr'k_{l}(r')-\int_{r_{l}^{(0)}}^{r}dr'k_{l}^{(0)}(r')\right], 
\end{equation}
where $k_{l}(r)$ and $k_{l}^{(0)}(r)$ are the local wave numbers in
the presence and absence of the potential, respectively,
\begin{equation}
\hbar k_{l}(r)= p_{l}(r)= \sqrt{2\mu\left(E-V(r)-\frac{\hbar^{2}(l+1/2)^{2}}{2\mu
    r^2}\right)} ,
\end{equation}
\begin{equation}
\hbar k_{l}^{(0)}(r)= p_{l}^{(0)}(r)=
\sqrt{2\mu\left(E-\frac{\hbar^{2}(l+1/2)^{2}}{2\mu r^2}\right)} ,
\end{equation}
where $\mu$, here, is the reduced mass. The radii, $r_{l}$ and $r_{l}^{(0)}$, are the classical turning points defined
by $p_{l}(r_{l}) = 0$, and $p_{l}^{(0)}(r_{l}^{(0)}) = 0$, respectively.
The phase shift, now a function of the energy (or the asymptotic wave
number $k$), and the semi-classical angular momentum, $\lambda = l+1/2$,
can be evaluated once the potential is given. In the case of
point-charge Coulomb scattering, the result of such a calculation
results in the following expression for the WKB phase shift function
$\delta(\lambda, k) = \sigma(\lambda, k)$,

\begin{equation}
\sigma(\lambda, k)= \frac{1}{2}\eta\ln\left[\eta^2 +\lambda^2\right] +
\lambda\sin^{-1}\left[\frac{\eta}{\sqrt{\eta^2 + \lambda^2}}\right] 
-\eta .
\label{delta-WKB}
\end{equation}
where the Sommerfeld parameter $\eta$ is related to the asymptotic wave number, $k$,  by, $\eta = ka$, where $a$ is half the distance of closest approach for head-on, zero impact parameter, collision.

The above expression should be compared to our
Eq.~(\ref{FormulaExataParaSigmal-B}). Clearly the major difference resides in 
the introduction of the semi-classical angular momentum variable
$\lambda$ in the former equation Eq. (\ref{delta-WKB}). This change is
in fact necessary when the WKB form of the wave function is invoked
\cite{Langer37}, to guarantee the presence of a classical turning point
for s-waves (now the angular momentum is not zero but rather $\hbar/2$). 

We turn next to the eikonal approximation. This, may be considered as
the high-energy limit of the WKB approximation. Expanding the WKB
phase shift in V/E, and keeping the linear term, and further assuming
$r^2 = z^2 + b^2$, with $b$ being the impact parameter, $kb =
\lambda$, we find for the eikonal Coulomb phase shift, $\sigma(b, k)$,
the following,
\begin{equation}
\sigma(b, k) = -\frac{\eta}{2} \int_{-\infty}^{\infty} dz \frac{1}{\sqrt{b^2 + z^2}}.
\end{equation}

The integral in the above equation diverges at both
extremes. Ref. \cite{Glauber59} introduced a screening function that
renders the integral finite. The screening function considered is
$F(r) = \Theta (a - r)$, where the screening length $a$ is of very
large, and taken to be $a \gg b$. Then,
\begin{equation}
\sigma(b,k) = 
-\frac{\eta}{2} 
\ln \left(\frac{2a + 2\sqrt{a^2 - b^2}}{2a - 2\sqrt{a^2 -b^2}}\right),
\end{equation}
which, with $a \gg b$, reduces to 
\begin{equation}
\sigma(b, k) = \eta \ln \left(\frac{b}{2a}\right) = \eta \ln \lambda - \eta \ln (2ka).
\end{equation}
The above form is similar to Eq.~(\ref{glauber}), except for the
change $l \rightarrow \lambda$ and the screening term, which depends
only on energy and not on $\lambda$.  Glauber \cite{Glauber59}
considered also and exponential screening function of the form $F(r) =
\exp{-r/a}$, and found the following limiting expression for the
Coulomb phase,
\begin{equation}
\sigma(b, k) = \eta \left[\ln \frac{b}{2a} - \gamma\right] ,
\end{equation}
where $\gamma$ is Euler's constant, $\gamma = 0.57721\ldots$. Again, the main
feature of Eq. (\ref{glauber}) of a logarithmic dependence on $b$, and
thus $\lambda$ is maintained. A Gaussian shaped screening function
$F(r) =\exp{[-r^2/a^2]}$ was considered by \cite{Hassan86}, and the
result found for the phase shift is,
\begin{equation}
\sigma(b, k) = \eta \left[\ln \frac{b}{2a} - \frac{\gamma}{2}\right].
\end{equation}
Once again the main feature of Eq. (\ref{glauber}) is
maintained, namely, the logarithmic dependence on b. Clearly the discussion above demonstrates that the best
route to follow to obtain well behaved and defined asymptotic forms
for the Coulomb phase shift is to rely on the exact expression,
Eq. (\ref{sigma-gamma}), and use the uniform approximation method of
Stirling's series to obtain the correct asymptotic form of the
$\Gamma$ function.\\

\section{Quantum and classical  Coulomb deflection functions}
An important theoretical entity which enters in any semi-classical treatment of scattering is the deflection
function. In the continuous $\lambda = l + 1/2$ limit, the deflection function is just the derivative of $2\sigma(\lambda, k)$ with respect to $\lambda$ \cite{Brink85}. Thus,

\begin{equation}
\Theta(\lambda) = 2\frac{d\sigma(\lambda, k)}{d\lambda}.
\end{equation}
Using the WKB expression, Eq. (\ref{delta-WKB}), we obtain the well known classical Rutherford deflection function,
\begin{equation}
\Theta(\lambda) = 2 \tan^{-1} \frac{\eta}{\lambda}.
\end{equation}
The above should be compared to the exact "quantum" deflection  function obtained from $\sigma_{l}$ of
Eq. (\ref{Formula-exata-para-sigma-l-I}), which can be re-written as a recursion formula, viz,
\begin{equation}
\sigma_{l} = \sigma_{l - 1} + \tan^{-1}\frac{\eta}{l}.
\end{equation}
Clearly one can define the derivative as merely the difference, $\sigma_{l} - \sigma_{l -1}$, and thus the
quantum deflection function, $\Theta_{q}(l)$ is,
\begin{equation}
\Theta_{q}(l) =  2 [\sigma_{l} - \sigma_{l - 1}] = 2 \tan^{-1}\frac{\eta}{l}.
\end{equation}
The quantum and classical deflection functions agree if  one makes the change $l \rightarrow\lambda$, as expected.\\

\section{Conclusions}
In this paper we have revisited the literature on the point Coulomb phase shift, and sharpened the
applicability  and range of validity of several of
the approximations usually employed to calculate it . In particular, the phase shift
at zero angular momentum, $\sigma_{0}(\eta)$,  as a function of the Sommerfeld  parameter, $\eta$, is calculated
exactly and a simple analytic expression for it valid for small and large $\eta$ is found. This is useful
when evaluating the phase shift as a function of angular momentum, $l$, using the exact formula,
 $\sigma_{l}(\eta) = \sigma_{0}(\eta) +\sum_{m=0}^{l}\tan^{-1}(\eta/m)$, of Eq. (\ref{Formula-exata-para-sigma-l-I}). \\

\textbf{Acknowledgments}\\
This work was supported in part by the CNPq, FAPESP and the MCT-National Institute of Quantum Information.\\

\end{document}